\documentclass[10pt,oneside,english]{elsart}
\usepackage[T1]{fontenc}
\usepackage[latin1]{inputenc}
\usepackage{array}
\usepackage{graphicx}
\usepackage{amssymb}

\makeatletter

\providecommand{\tabularnewline}{\\}

\usepackage{babel}
\makeatother
\begin{document}
\begin{frontmatter}

\title{Characterization of welding defects by fractal analysis of ultrasonic
signals}

\author{A. P. Vieira, E. P. de Moura, L. L. Gonçalves}

\address{Departamento de Engenharia Metalúrgica e de Materiais, Universidade
Federal do Ceará, Fortaleza, CE, Brazil}

\author{J. M. A. Rebello}

\address{Departamento de Engenharia Metalúrgica e de Materiais, Universidade
Federal do Rio de Janeiro, RJ, Brazil}

\begin{abstract}
In this work we apply tools developed for the study of fractal properties
of time series to the problem of classifying defects in welding joints
probed by ultrasonic tecniques. We employ the fractal tools in a preprocessing
step, producing curves with a considerably smaller number of points
than in the original signals. These curves are then used in the classification
step, which is realized by applying an extension of the Karhunen-Loève
linear transformation. We show that our approach leads to small error
rates, comparable with those obtained by using more time-consuming
methods based on non-linear classifiers.
\end{abstract}
\end{frontmatter}

\section{Introduction}

Ultrasonic tests can serve as a useful tool for evaluating the integrity
of metallic structures, and specially of weld joints. By inspecting
the scattering pattern of ultrasonic waves propagating in the material,
it is possible to identify the presence of defects, and to estimate
their dimensions. However, it is often desirable to have precise information
about the nature of the defects, and a number of studies have tried
to propose useful approaches to perform such classification \cite{masnata96,margrave99,moura05a,moura05b},
mostly based on direct analysis of the patterns with neural networks. 

In the present paper, we describe a distinct approach, based on tools
developed for analyzing fractal properties of time series \cite{hurst51,peng94,dubovikov04,addison97}.
Such kind of approach has been successfully applied to ultrasonic
signals (interpreted as a particular kind of time series) by a number
of authors \cite{barat98,matos04,silva05}, both in defect and in
microstructure classification, by calculating the exponents of the
power laws characterizing various fractal features of the series,
and hoping to associate different sets of exponents with different
classes. However, this can only be expected to work when the typical
series corresponding to different classes are highly dissimilar, due
to the fact that the size of each series is usually small, and estimates
of the various exponents are subject to significant fluctuations.
Thus, in general, an expanded set of features must be used to obtain
an efficient classification algorithm. Here we employ tools from the
statistical pattern-classification literature to extract relevant
features from the set of fractal analyses applied to ultrasonic signals
obtained from weld joints having three different kinds of defects.

\begin{figure}
\begin{centering}\includegraphics[width=0.8\columnwidth]{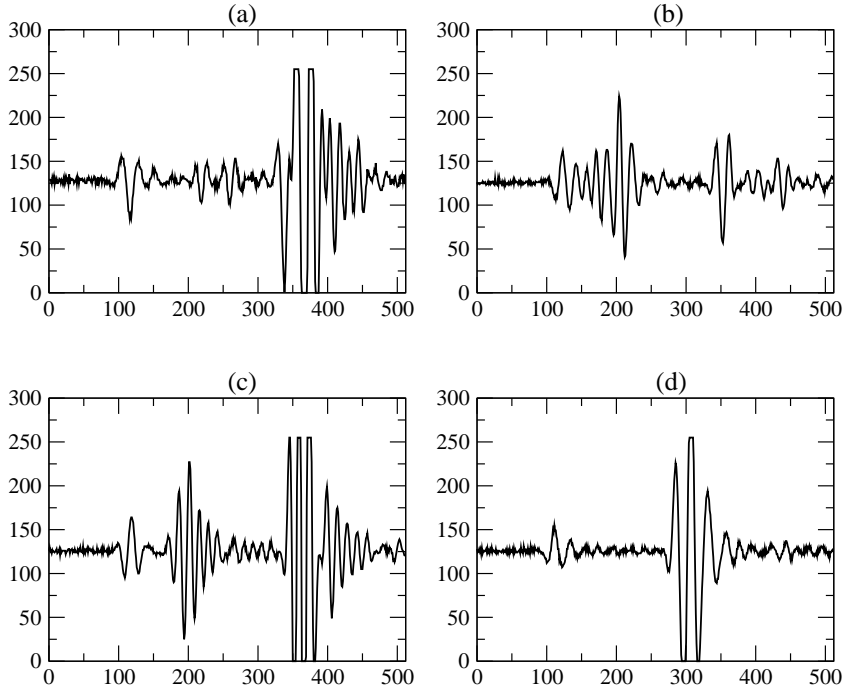}\par\end{centering}

\caption{\label{fig:exemplosdesinais}Typical examples of signals obtained
from samples with (a) lack-of-fusion defects, (b) lack-of-penetration
defects, (c) porosities, and (d) no defects. The horizontal axes correspond
to the time direction, in units of the inverse sample rate of the
equipment.}
\end{figure}

We used $240$ ultrasonic signals obtained by the TOFD tecnique \cite{silk76},
with $60$ signals corresponding to each kind of defect (lack of fusion,
lack of penetration, and porosity) and other $60$ signals from regions
with no defects. (For a description of the materials used, as well
as tecniques for producing and capturing the signals, see Ref. \cite{moura05a}.)
All signals had a length of $512$ points, with $8$-bit resolution.
Typical signals are shown in Fig. \ref{fig:exemplosdesinais}. After
normalizing all signals so that the maximum and minimum values correspond
to $1$ and $-1$, we calculated the corresponding curves from four
different techniques of fractal analysis, which we describe in Sec.
\ref{sec:Fractal-analyses}. Then, as described in Sec. \ref{sec:Results-and-discussion},
we employed a variation of the Karhunen-Loève (KT) linear transformation
\cite{kittler73,webb02} to extract relevant features from the curves.
As we discuss in the final section, the combined approach of fractal
analysis and KT transformation yields a quite good classification
tool for the defects studied. %

\section{Fractal analysis}

\label{sec:Fractal-analyses}All techniques of fractal analysis employed
here start by dividing the signal into intervals containing $\tau$
points. Each technique then involves the calculation of the average
of some quantity $Q\left(\tau\right)$ over all intervals, for different
values of $\tau$. In a signal with genuine fractal features, $Q\left(\tau\right)$
should scale as a power of $\tau$, \begin{equation}
Q\left(\tau\right)\sim\tau^{\eta},\end{equation}
at least in an intermediate interval of values of $\tau$, corresponding
to $1\ll\tau\ll L$, $L$ being the signal length. 

Fractals of different nature should give rise to different exponents
$\eta$, providing a signature of the fractal. In our case, due to
the finite amount of points, and to the very nature of the signals,
a pure power-law behavior is hard to observe. Instead, as shown in
Fig. \ref{fig:curvasFF}, the curves usually exhibit features such
as a crossover between different power-law behaviors, or saturation
points, which can also serve as signatures of the different kinds
of defects. However, identifying the relevant features in advance
is a complex task. Fortunately, the pattern-classification literature
offers useful tools for feature extraction from data, and we describe
one of those in Sec. \ref{sec:Results-and-discussion} and Appendix
\ref{sec:Karhunen-Lo=E8ve-transformation}.

\subsection{Hurst (R/S) analysis}

The rescaled-range (R/S) analysis was introduced by Hurst \cite{hurst51}
as a tool for evaluating the persistency or antipersistency of a time
series. The method works by dividing the series into intervals of
a given size, and calculating the average ratio of the range (the
difference between the maximum and minimum values of the series) to
the standard deviation inside each interval. The size of each interval
is then varied.%
\begin{figure}
\begin{centering}\includegraphics[width=0.8\columnwidth]{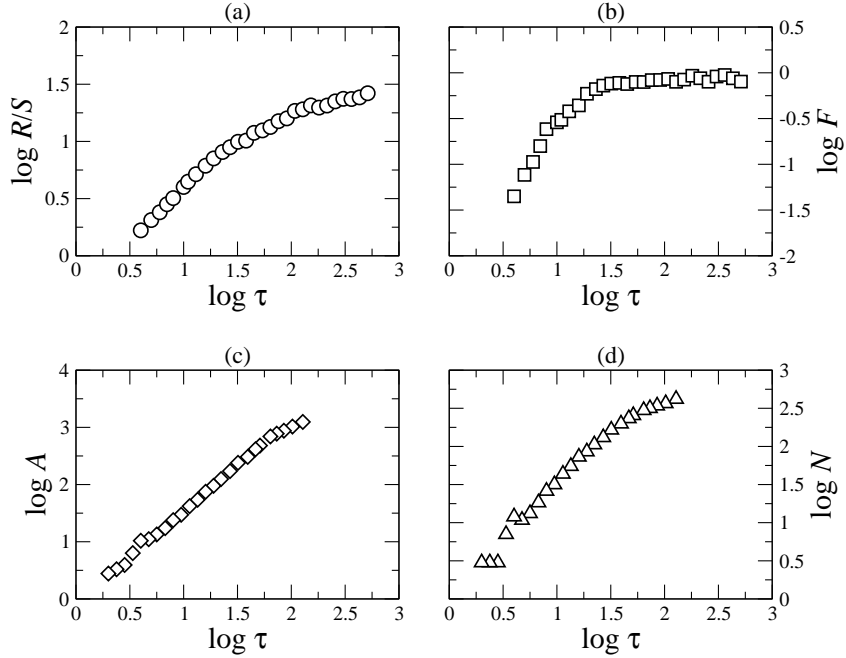}\par\end{centering}

\caption{\label{fig:curvasFF}Curves for a lack-of-fusion signal, obtained
from (a) Hurst analysis, (b) detrended-fluctuation analysis, (c) minimal-cover
analysis, and (d) box-counting analysis.}
\end{figure}

Mathematically, the R/S analysis is defined in the following way.
Given an interval of size $\tau$, whose left end is located at point
$i_{0}$, we calculate $\langle z\rangle_{\tau}$, the average of
the series $z_{i}$ inside the interval, \begin{equation}
\langle z\rangle_{\tau}=\frac{1}{\tau}\sum_{i=i_{0}}^{i_{0}+\tau-1}z_{i}.\end{equation}
 We then define an accumulated deviation from the mean as \begin{equation}
Z_{i}=\sum_{k=i_{0}}^{i}\left(z_{k}-\langle z\rangle_{\tau}\right),\end{equation}
 from which we extract a range, \begin{equation}
R(\tau)=\max_{i_{0}\leqslant i\leqslant i_{0}+\tau-1}Z_{i}-\min_{i_{0}\leqslant i\leqslant i_{0}+\tau-1}Z_{i},\end{equation}
 and the corresponding standard deviation, \begin{equation}
S(\tau)=\sqrt{\frac{1}{\tau}\sum_{i=i_{0}}^{i_{0}+\tau-1}Z_{i}^{2}}.\end{equation}
Finally, we obtain the rescaled range $R(\tau)/S(\tau)$, and take
its average over all intervals.

For a curve with true fractal features, the rescaled range should
satisfy the scaling form \begin{equation}
\frac{R(\tau)}{S(\tau)}\thicksim\tau^{H},\end{equation}
where $H$ is the Hurst exponent.

A typical curve obtained from the R/S analysis of the signals is shown
in Fig. \ref{fig:curvasFF}(a).

\subsection{Detrended-fluctuation analysis}

The detrended-fluctuation analysis (DFA) \cite{peng94} aims to improve
the evaluation of correlations in a time series by eliminating trends
in the data.

The method consists initially in obtaining a new integrated series
$\tilde{z}_{i}$, \begin{equation}
\tilde{z}_{i}=\sum_{k=1}^{i}\left(z_{k}-\langle z\rangle\right),\end{equation}
the average $\langle z\rangle$ being taken over all points, \begin{equation}
\langle z\rangle=\frac{1}{L}\sum_{i=1}^{L}z_{i}.\end{equation}
After dividing the series into intervals, the points inside a given
interval are fitted by a polynomial curve of degree $n$. In our case,
we have considered $n=1$ or $n=2$, corresponding to first- and second-order
fits. Then, a detrended variation function $\Delta_{i,n}$ is obtained
by subtracting from the integrated data the local trend as given by
the fit. Explicitly, we define \begin{equation}
\Delta_{i,n}=\tilde{z}_{i}-h_{i,n},\end{equation}
where $h_{i,n}$ is the value associated with point $i$ according
to the fit of degree $n$. Finally, we calculate the root-mean-square
fluctuation $F_{n}(\tau)$ inside an interval as \begin{equation}
F_{n}(\tau)=\sqrt{\frac{1}{\tau}\sum_{i}\Delta_{i,n}^{2}},\end{equation}
and average over all intervals. For a true fractal curve, $F(\tau)$
should behave as \begin{equation}
F(\tau)\thicksim\tau^{\alpha},\end{equation}
where $\alpha$ is the scaling exponent.

A typical curve obtained from the detrented-fluctuation analysis of
the signals is shown in Fig. \ref{fig:curvasFF}(b).

\subsection{Minimal-cover analysis}

This recently introduced method \cite{dubovikov04} relies on the
calculation of the minimal area necessary to cover a given plane curve
at a specified scale.

After dividing the series, we can associate with each interval, labeled
by a variable $k$, a rectangle of height $H_{k}$, defined as the
difference between the maximum and minimum values of the series $z_{i}$
inside the $k$th interval, \begin{equation}
H_{k}=\max_{i_{0}\leqslant i\leqslant i_{0}+\tau-1}z_{i}-\min_{i_{0}\leqslant i\leqslant i_{0}+\tau-1}z_{i},\end{equation}
in which $i_{0}=1+\left(k-1\right)\tau$ labels the left end of the
interval. The minimal area is then given by \begin{equation}
A\left(\tau\right)=\sum_{k}H_{k}\tau,\end{equation}
the summation running over all cells.

Ideally, in the scaling region, $A(\tau)$ should behave as \begin{equation}
A\left(\tau\right)\thicksim\tau^{2-D_{\mu}},\end{equation}
where $D_{\mu}$ is the minimal cover dimension, which is equal to
$1$ when the signal presents no fractality.

A typical curve obtained from the minimal-cover analysis of a signal
is shown in Fig. \ref{fig:curvasFF}(c).

\subsection{Box-counting analysis}

This is a well-know method of estimating the fractal dimension of
a point set \cite{addison97}, and it works by counting the minimum
number $N\left(\tau\right)$ of boxes of side $\tau$ needed to cover
all points in the set. For a real fractal, $N\left(\tau\right)$ should
follow a power law whose exponent is the box-counting dimension $D_{B}$,\begin{equation}
N\left(\tau\right)\sim\tau^{-D_{B}}.\end{equation}

A typical box-counting curve for a signal is shown in Fig. \ref{fig:curvasFF}(d).

\section{Results of the classification approach}

\label{sec:Results-and-discussion}
\begin{figure}
\begin{centering}\includegraphics[width=0.7\columnwidth]{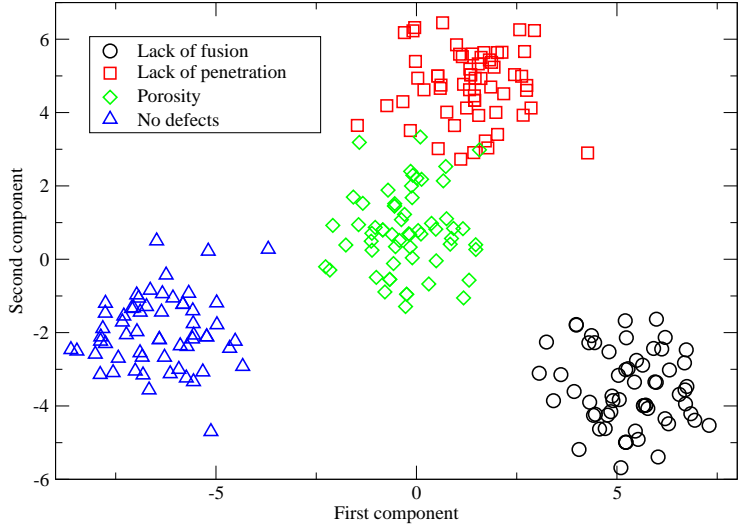}\par\end{centering}

\caption{\label{fig:KLUU12todos}Projection, along the first two components,
of the vectors obtained by applying the Karhunen-Loève transformation
to the to the full data set obtained from four different fractal analyses.}
\end{figure}

In order to classify the signals,
we used a supervised variation of the Karhunen-Lo\`{e}ve (KL) transformation
\cite{kittler73,webb02}, briefly described in Appendix \ref{sec:Karhunen-Lo=E8ve-transformation}.
For each signal, we collected the corresponding curves from various
fractal analyses, forming a single vector with $M$ components. The
most successfull combination involves curves from Hurst, linear detrended-fluctuation,
minimal-cover, and box-counting analyses, corresponding to $M=108$
(with $27$ components of the vector taken from each curve). A plot
obtained by projecting the first two components of the KL-transformed
vectors is shown in Fig. \ref{fig:KLUU12todos}, for the full set
of vectors. (Note that, with $4$ different classes for the vectors,
the transformed space is three-dimensional.) It is evident from the
figure that the transformation yields a good clustering of the vectors
around the different class means. This clustering is a general feature
of the KL transformation. However, to assess the utility of the classification
approach, it is essential to evaluate the generalization error.%

We proceeded by first randomly dividing the vectors into a training
set (with $80\%$ of the signals) and a test set (with the remaining
signals). The KL transformation was first applied to the training
vectors, and the class means were determined. Transformed vectors
in both sets were then classified by applying the nearest-class-mean
rule, i.e., a vector $\mathbf{x}$ was assigned to the class whose
average vector, as determined by the training set, lies closer to
$\mathbf{x}$. (It is also possible to explore other approaches for
discrimination, such as Bayesian rules , but that would require an
estimation of the class-conditional probabilities, which we do not
have at hand.) Finally, we took averages over $500$ different choices
of training and test sets. 

The average confusion matrices of the training and test sets are shown
in Tables \ref{tab:1} and \ref{tab:2}. Notice that the mean error
rate is negligible for the training vectors, and corresponds to around
$15\%$ for the test vectors. These error rates are comparable to
those obtained by analyzing the same signals directly using non-linear
classifiers based on neural networks \cite{moura05b}; the use of
linear classifiers, on the other hand, leads to considerably higher
error rates \cite{moura05a}. Notice that in our study the number
of variables ($108$) employed in the classification step represents
only around $1/5$ of the number used in the neural-network studies
(which made use of all $512$ points of each signal). Besides rendering
the calculations faster, for an equivalent error rate, the smaller
number of variables also leads to smaller fluctuations in the curves.%
\begin{table}

\caption{Average confusion matrix for the training vectors, derived from the
fractal analyses. The possible classes are lack of fusion (LF), lack
of penetration (LP), porosity (PO) and no defects (ND). The figures
in parenthesis indicate the standard deviations, calculated over $500$
sets. The value in row $i$, column $j$ indicates the percentage
of vectors belonging to class $i$ which were associated with class
$j$.\label{tab:1}}

\noindent \begin{centering}\begin{tabular}{|c|>{\centering}p{0.16\columnwidth}|>{\centering}p{0.16\columnwidth}|>{\centering}p{0.16\columnwidth}|>{\centering}p{0.16\columnwidth}|}
\hline 
&
LF &
LP &
PO &
ND \tabularnewline
\hline 
LF &
\textbf{100 }&
0&
0 &
0 \tabularnewline
\hline 
LP &
0 &
\textbf{99.87} (0.50)&
0.13 (0.50) &
0 \tabularnewline
\hline 
PO &
0 &
0.01 (0.10) &
\textbf{99.99} (0.10) &
0\tabularnewline
\hline 
ND &
0 &
0&
0.01 (0.08) &
\textbf{99.99} (0.08) \tabularnewline
\hline
\end{tabular}\par\end{centering}
\end{table}
\begin{table}

\caption{The same as in Table \ref{tab:1}, for the testing vectors.\label{tab:2}}

\noindent \begin{centering}\begin{tabular}{|c|>{\centering}p{0.165\columnwidth}|>{\centering}p{0.165\columnwidth}|>{\centering}p{0.165\columnwidth}|>{\centering}p{0.165\columnwidth}|}
\hline 
&
LF &
LP &
PO &
ND \tabularnewline
\hline 
LF &
\textbf{91.07} (8.20) &
1.69 (3.64)&
6.88 (7.37) &
0.35 (1.69) \tabularnewline
\hline 
LP &
2.61 (8.20) &
\textbf{83.96} (10.04)&
12.14 (9.27) &
1.28 (3.21) \tabularnewline
\hline 
PO &
6.43 (7.27) &
13.99 (10.5) &
\textbf{72.66} (12.87)&
6.92 (7.55)\tabularnewline
\hline 
ND &
1.01 (3.25) &
2.55 (4.43)&
6.92 (7.16) &
\textbf{89.51} (8.93) \tabularnewline
\hline
\end{tabular}\par\end{centering}
\end{table}

For completeness, we also applied the KT transformation to both the
correlograms and the Fourier spectra of each signal, obtaining average
error rates no smaller than $36\%$ and $48\%$, respectively (see
Table \ref{tab:3}). %
\begin{table}

\caption{\label{tab:3}Average percentage rates of correct classification
of the test vectors, derived from applying the KL transformation to
the correlograms or the Fourier spectra associated with the signals.
The figures in parenthesis indicate standard deviations calculated
over $100$ sets.}

\noindent \begin{centering}\begin{tabular}{|c|c|c|c|c|}
\hline 
&
LF&
LP&
PO&
ND\tabularnewline
\hline
\hline 
Correlograms&
60.19 (17.75)&
64.18 (14.96)&
51.60 (16.87)&
57.43 (16.16)\tabularnewline
\hline 
Fourier spectra&
51.42 (17.49)&
47.35 (17.03)&
46.76 (16.64)&
48.08 (15.76)\tabularnewline
\hline
\end{tabular}\par\end{centering}
\end{table}

\section{Conclusions}

In this paper we applied techniques developed for the study of fractal
properties of time series as a preprocessing tool for the classification
of defects probed by ultrasonic signals. The signals were obtained
in welding joints containing three different classes of defects, and
we also considered signals with no defects. For the classification
step, we employed an extension of the Karhunen-Loève transformation,
which, supplemented by the nearest-class-mean rule, yielded low error
rates (between $0$ and $15\%$) both in the training and the test
stages. These error rates are comparable with those obtained from
more time-consuming approaches based on direct analysis of the signals.
In our view, this is evidence that fractal techniques are a promising
tool for the classification of defects probed by ultrasonic inspection.

We believe that the performance of the classification approach based
on fractal techniques can be further improved by resorting to non-linear
classifiers, especially in combination with reclassification and hierarchical
procedures. These extensions we leave for future investigations.

\begin{ack}
We acknowledge financial support from the Brazilian agencies FUNCAP,
CNPq, CAPES and FINEP (CT-Petro). A. P. Vieira has benefitted from
helpful conversations with N. Caticha.
\end{ack}
\appendix

\section{Karhunen-Loève transformation\label{sec:Karhunen-Lo=E8ve-transformation}}

The Karhunen-Loève (KL) transformation, as the principal component
analysis, is a tool for feature selection and extraction. It produces
a set of mutually uncorrelated components, and dimensionality reduction
can be achieved by selecting those components with the largest variances.
The version of the transformation employed here {[}Kittler and Young,
1973 (Webb)] relies on compression of the discriminatory information
contained in the class means.

Let $\mathbf{x}_{i}$ be the (column) vector corresponding to the
$i$th signal. The KL transformation consists of first projecting
the training vectors along the eigenvectors of the within-class covariance
matrix $\mathbf{S}_{W}$, defined by \begin{equation}
\mathbf{S}_{W}=\frac{1}{N}\sum_{k=1}^{N_{C}}\sum_{i=1}^{N_{k}}y_{ik}(\mathbf{x}_{i}-\mathbf{m}_{k})(\mathbf{x}_{i}-\mathbf{m}_{k})^{T},\end{equation}
where $N_{C}$ is the number of different classes, $N_{k}$ is the
number of vectors in class $k$, $\mathbf{m}_{k}$ is the average
vector of class $k$, and $T$ denotes the transpose of a matrix (in
this case, yielding a row vector). The element $y_{ik}$ is equal
to one if $\mathbf{x}_{i}$ belongs to class $k$, and zero otherwise.
We also rescale the resulting vectors by a diagonal matrix built from
the eigenvalues $\lambda_{j}$ of $\mathbf{S}_{W}$. In matrix notation,
this operation can be written as \begin{equation}
\mathbf{X}^{\prime}=\Lambda^{-\frac{1}{2}}\mathbf{U}^{T}\mathbf{X},\end{equation}
in which $\mathbf{X}$ is the matrix whose columns are the training
vectors $\mathbf{x}_{i}$, $\Lambda=\mathrm{diag}(\lambda_{1},\lambda_{2},...)$,
and $\mathbf{U}$ is the matrix whose columns are the eigenvectors
of $\mathbf{S}_{W}$. This choice of coordinates makes sure that the
transformed within-class covariance matrix corresponds to the unit
matrix. Finally, in order to compress the class information, we project
the resulting vectors onto the eigenvectors of the between-class covariance
matrix $\mathbf{S}_{B}$, \begin{equation}
\mathbf{S}_{B}=\sum_{k=1}^{N_{C}}\frac{N_{k}}{N}(\mathbf{m}_{k}-\mathbf{m})(\mathbf{m}_{k}-\mathbf{m})^{T},\end{equation}
where $\mathbf{m}$ is the overall average vector. The full transformation
can be written as \begin{equation}
\mathbf{X}^{\prime\prime}=\mathbf{V}^{T}\Lambda^{-\frac{1}{2}}\mathbf{U}^{T}\mathbf{X},\end{equation}
$\mathbf{V}$ being the matrix whose columns are the eigenvectors
of $\mathbf{S}_{B}$ (calculated from $\mathbf{X}^{\prime}$).

With $N_{C}$ possible classes, the fully-transformed vectors have
at most $N_{C}-1$ relevant components. We then associate a vector
$\mathbf{x}_{i}$ with the class whose average vector lies closer
to $\mathbf{x}_{i}$ within the transformed $(N_{C}-1)$-dimensional
space. This association rule would be optimal if the vectors in different
classes followed normal distributions.

\bibliographystyle{unsrt}
\bibliography{end}

\begin{thebibliography}{10}

\bibitem{masnata96}
A.~Masnata and M.~Sunseri.
\newblock {\em NDT\&E International}, 29:87--93, 1996.

\bibitem{margrave99}
F.~W. Margrave, K.~Rigas, D.~A. Bradley, and P.~Barrowcliffe.
\newblock {\em Measurement}, 25:143--154, 1999.

\bibitem{moura05a}
E.~P. {de Moura}, M.~H.~S. Siqueira, R.~R. {da Silva}, J.~M.~A. Rebello, and
  L.~P. Cal\^oba.
\newblock {\em Insight}, 47:777--782, 2005.

\bibitem{moura05b}
E.~P. {de Moura}, M.~H.~S. Siqueira, R.~R. {da Silva}, and J.~M.~A. Rebello.
\newblock {\em Insight}, 47:783--787, 2005.

\bibitem{hurst51}
H.~E. Hurst.
\newblock {\em Trans. Am. Soc. Civ. Eng.}, 116:770--799, 1951.

\bibitem{peng94}
C.~K. Peng, V.~Buldyrev, S.~Havlin, M.~Simmons, H.~E. Stanley, and A.~L.
  Goldberger.
\newblock {\em Phys. Rev. E}, 49:1685--1689, 1994.

\bibitem{dubovikov04}
P.~M. Dubovikov, N.~V. Starchenko, and M.~S. Dubovikov.
\newblock {\em Physica A}, 339:591--608, 2004.

\bibitem{addison97}
P.~S. Addison.
\newblock {\em Fractals and Chaos}.
\newblock IOP, London, 1997.

\bibitem{barat98}
P.~Barat.
\newblock {\em Chaos, Solitons \& Fractals}, 9:1827--1834, 2004.

\bibitem{matos04}
J.~M.~O. Matos, E.~P. {de Moura}, S.~E. Kr\"uger, and J.~M.~A. Rebello.
\newblock {\em Chaos, Solitons \& Fractals}, 19:55--60, 2004.

\bibitem{silva05}
F.~E. Silva, L.~L. Gon\c{c}alves, D.~B.~B. Ferreira, and J.~M.~A. Rebello.
\newblock {\em Chaos, Solitons \& Fractals}, 26:481--494, 2005.

\bibitem{silk76}
M.~G. Silk.
\newblock In R.~S. Sharpe, editor, {\em Research Techniques in NDT}, volume
  III, page~51. Academic Press, London, 1976.

\bibitem{kittler73}
J.~Kittler and P.~C. Young.
\newblock {\em Pattern Recognition}, 5:335--352, 1973.

\bibitem{webb02}
A.~R. Webb.
\newblock {\em Statistical Pattern Recognition, 2nd ed.}
\newblock John Wiley \& Sons, West Sussex, 2002.

\end{thebibliography}

\end{document}